# Gold Nanoparticles Coated Optical Fiber for Real-time Localized Surface Plasmon Resonance Analysis of *In-situ* Light-Matter Interactions

*Nafize Ishtiaque Hossain[1,*], Kazi Zihan Hossain[2,*], Momena Monwar[2], Md. Shihabuzzaman Apon[3,↓], Caleb Shaw[2,↓], Shoeb Ahmed[3], Shawana Tabassum[1,#], M. Rashed Khan[2,#].*

[1]Department of Electrical Engineering, University of Texas, Tyler; Texas, U.S.A.

[2]Department of Chemical and Materials Engineering, University of Nevada, Reno; Nevada, U.S.A.

[3]Department of Chemical Engineering, Bangladesh University of Engineering and Technology; Dhaka, Bangladesh.

*,↓Contributed equally

[#]Corresponding Authors: Shawana Tabassum: stabassum@uttyler.edu

M. Rashed Khan mrkhan@unr.edu

**Abstract:**

*In situ* measurement of analytes for *in vivo* or *in vitro* systems has been challenging due to the bulky size of traditional analytical instruments. Also, frequent *in vitro* concentration measurements rely on fluorescence-based methods or direct slicing of the matrix for analyses. These traditional approaches become unreliable if localized and *in situ* analyses are needed. In contrast, for *in situ* and real-time analysis of target analytes, surface-engineered optical fibers can be leveraged as a powerful miniaturized tool, which has shown promise from bio to environmental studies. Herein, we demonstrate an optical fiber functionalized with gold nanoparticles using a dip-coating process to investigate the interaction of light with molecules at or near the surface of the optical fiber. Localized surface plasmon resonance from the light-matter interaction enables the detection of minute changes in the refractive index of the surrounding medium. We used this principle to assess the *in situ* molecular distribution of a synthetic drug (methylene blue) in an *in vitro* matrix (agarose gel) having varying concentrations. Leveraging the probed Z-height in diffused analytes, combined with its *in silico* data, our platform shows the feasibility of a simple optofluidic tool. Such straightforward *in situ* measurements of analytes with optical fiber hold potential for real-time molecular diffusion and molecular perturbation analyses relevant to biomedical and clinical studies.

## Introduction

*In situ* analysis of biomolecules with high-throughput spatiotemporal fidelity is desired in various applications, including imaging and sensing, healthcare, environmental monitoring, and precision farming.[1–3] Spectro-analytical systems (i.e., LC/GC-MS, UV-VIS, MALDI, DESI) are bulky, expensive, and lack *in situ* monitoring capabilities; however, these are powerful analytical tools for *ex situ* and *in vitro* cases. In contrast, *in situ* and real-time analysis of target analytes distributed in a three-dimensional (3D) biomatrix frequently require miniaturized and inexpensive analytical tools. For example, *in vitro* molecular diffusion analysis of a synthetic drug primarily relies on image-based or fluorescence-based microscopic techniques;[4] however, such techniques face challenges if the matrix becomes opaque or distribution is beyond the imaging limit (i.e., a few layers below the top surface). In such instances, miniaturized tools that provide z-depth profiling can become very powerful. Electroanalytical tools,[5,6] capacitive probes,[7] dialysis probes[8] or microneedles-based systems[9] offer volumetric sample analysis within a biomatrix. However, electrode preparation for electroanalytical/microneedle-based experiments or membrane preparation in dialysis-based experiments are time-limiting and require specialized clean-room-based metal electrode or membrane preparation techniques.[10,11] Concurrently, biometrics are frequently sliced to avoid the complexities of analyzing the distribution of an analyte in 3D.[12] These traditional microscopic approaches for *in situ* concentration measurements are briefly summarized in **Figure 1a**. In contrast, easily insertable miniaturized spectrophotometric analytical devices for *in vivo* or *in vitro* analysis, harnessing the interactions of light with the diffusing fluid and materials (i.e., glass, polymer) can provide a new platform to sense the local concentration. By adjusting the insertion depth, this approach can also allow for investigating the concentration at various depths. These z-depth investigations seem challenging and impossible to execute using bulky spectroanalytical, electroanalytical, or membrane-based dialysis systems. In this study, we report a metal nanoparticle-coated optical fiber tip and its feasibility in investigating the z-depth profiling of methylene blue (analyte) diffused into agarose (biomatrix).

Our functionalized fiber-optic surface was developed through a simple dip-coating procedure without using any cleanroom-based fabrication technologies. This is in contrast to intricately engineered micromachined fibers requiring costly and advanced lithography techniques for *in situ* optofluidic analysis.[13–15] Furthermore, integrating micromachined fibers into fluidic systems requires complex alignment, which often decreases the signal yields of the final device.[16] Our approach facilitates simplicity in fabrication, improved tunability, and provides *in situ* and real-time analysis capabilities. Analyzing surface adsorption with such fiber optics has shown promise and versatility in various fields, from biotechnology to environmental monitoring.[17–19] The key principle behind this method is the interaction of light with molecules at or near the surface of the optical fiber, leading to changes in the optical properties that can be precisely measured.[20,21] This change in optical properties enables the real-time monitoring of molecular adsorption processes at interfaces. Furthermore, fiber optics offer advantages in terms of miniaturization and flexibility. The small diameter and flexibility of optical fibers make them suitable for implantable sensors or reaching confined spaces that traditional sensors cannot access, providing the opportunity to study *in situ* light-matter interactions. In this work, we have harnessed the light-matter interactions in a gold nanoparticles (AuNP) coated optical fiber to probe the *in vitro* concentration of a drug mimic.

We have used methylene blue, M.B. as the synthetic drug,[22] and agarose gels as the synthetic mimic of the brain matrix. Agarose has been widely used in the literature for mimicking the properties of brain tissue *in vitro* studies.[23–26] Healthy brain tissue can be simulated using a 0.6% (%w/v) agarose gel, while 2% (%w/v) agarose can represent unhealthy tissue.[24,27,28] Various agarose matrices with different concentrations provide the flexibility to optimize drug delivery scenarios before clinical trials. However, assessing the 3D distribution of a solute within a polymeric matrix in real-time, without damaging the sample, is challenging. It is essential to capture images and extract concentration profiles based on image intensity to investigate the instantaneous diffusion profile.[29] Without fluorescent dyes and a suitable camera with an automated system, obtaining an accurate 3D diffusion profile is not feasible. Simultaneously administering synthetic drugs and employing the image acquisition technique presents an additional challenge. Techniques like microdialysis demand an extended collection period, typically around 30 minutes, eliminating the instantaneous sampling factor. In contrast, we show a more straightforward solution to this problem by incorporating a fiber-optic probe that can be conveniently used in conjunction with the diffusion probe.

Our metal-functionalized optical fiber exhibited localized surface plasmon resonance (LSPR).[30–32] At the LSPR wavelength, the interaction of light with the synthetic drug, M.B. produced a reflected light intensity proportional to the concentration of the drug in deionized (DI) water as well as in an agarose gel matrix. Previously in the literature, silver nanostructures,[33] silver nanoparticle-reduced graphene oxide,[34] gold-titanium oxide yolk-shell nanostructure,[35] and sodium-tungsten-bronze nanosheets[36] have been used to detect M.B. In this present work, we have used AuNP-coated optical fiber and altered the position of the fiber to gather instantaneous 3D diffusion data of M.B., a novel contribution not yet found in existing literature. We placed the fiber probe at four Z-depths in three different hydrogel concentrations to investigate the molecular interactions of the synthetic drug with light. Additionally, we observed a good reproducibility of the fiber-optic response in different concentrations of the gel medium (1%, 2%, and 5%). We further compared the measured concentration with *in silico* profiles to gain confidence in our data.

We have presented our combined approach in **Figure 1b**, contrasting the traditional microscopic techniques in **Figure 1a.** Our proposed optical-fiber-based *in situ* measurement and *in silico* validation can be helpful for biomedical and bioengineering studies having clinical significance.

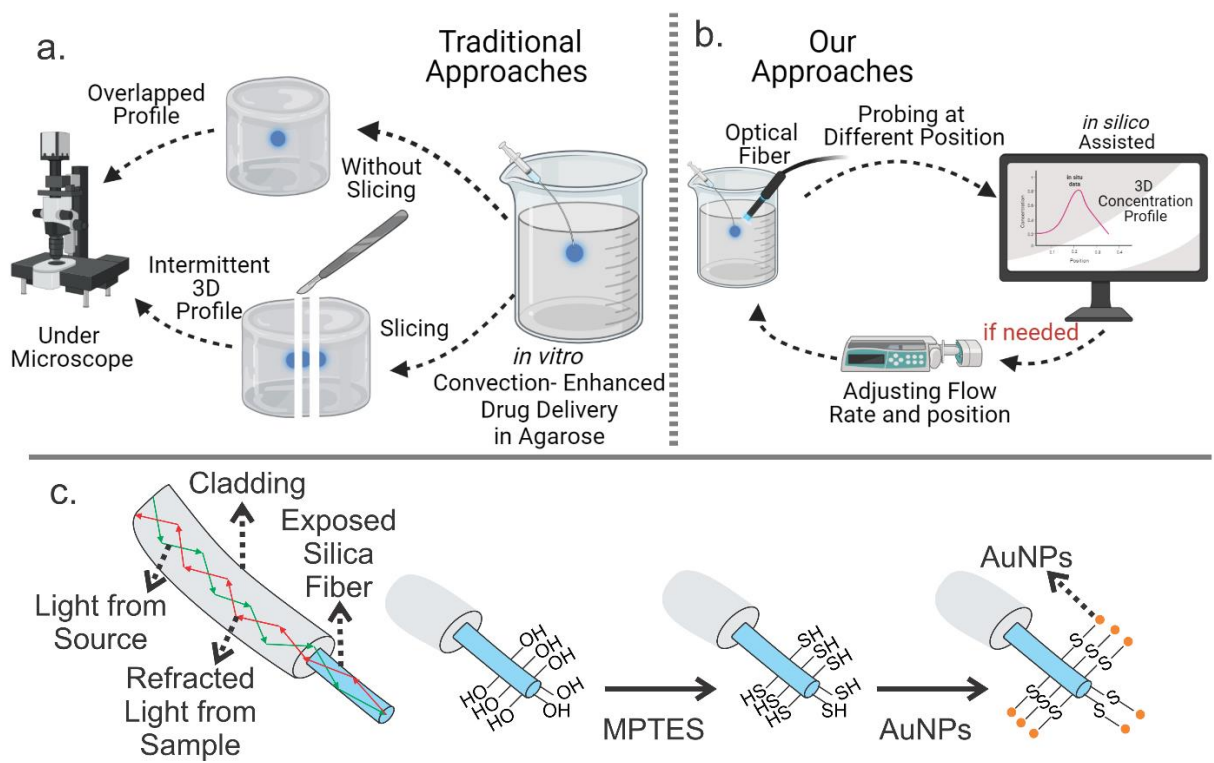

**Figure 1:** (a) Traditional approach for *in situ* drug concentration measurement for *in vitro* convection-enhanced drug delivery, lacking proper 3D information. (b) Our proposed approach will aid the process by measuring the local concentration with optical fiber, which will be assisted with *in silico* profiles for complete 3D information. (c) the workflow to make an AuNP-coated functional optical fiber.

## Materials and Methods

High-quality laboratory-grade chemicals were utilized without the need for additional purification. $HAuCl_4$ (Hydrogen tetrachloroaurate (III) hydrate), trisodium citrate, $H_2SO_4$ (sulfuric acid), HCl

(hydrochloric acid), MPTES (3-mercaptopropyltriethoxysilane), methanol, and acetic acid were purchased from Sigma Aldrich (St. Louis, MO). All experiments were conducted using deionized (DI) water with a resistivity of ~18 MΩ.

**Fabrication of the Fiber-tip LSPR Probe:** A section of approximately 1 cm at the end of a 200 µm diameter fused fiber was prepared for functionalization with gold nanoparticles (AuNPs) using a gold-thiol chemistry process. **Figure 1c** represents the gold-thiol chemistry on the fiber surface that we have used in this work. The initial step involved removing the cladding from the silica fiber using acetone. Then, the exposed portion of the fiber was cleaned and treated with Piranha solution (a mixture of $H_2SO_4$ and $H_2O_2$) at a 7:3 volume ratio for 30 minutes at 85ºC, which created hydroxyl (-OH) groups on the fiber surface. Subsequently, the optical fiber was rinsed with deionized (DI) water, dried with nitrogen gas, and annealed in a vacuum oven at 110ºC for 30 minutes.

The fiber surface was then modified with MPTES to introduce thiol (-SH) groups. This involved a sol-gel deposition process using an aqueous/alcohol solution consisting of 75% distilled water and 25% methanol, adjusted to a pH of approximately 4.5 with acetic acid.[30–32] MPTES was added to achieve a final concentration of 2% (v/v), and the solution was allowed to react for 10 minutes to facilitate alkoxide hydrolysis and silanol formation. The fiber tip was immersed in this solution for 30 minutes and subsequently rinsed with ethanol to remove any unbound silane. This step eliminated excess MPTES and unreacted species from the fiber surface. The fiber was then annealed at 110ºC for 30 minutes to promote condensation reactions.

The thiol-functionalized fiber was incubated overnight in a solution containing AuNPs, which were prepared through the citrate reduction of $HAuCl_4$, following the Turkevich method.[37] For 0.01% (w/v) of $HAuCl_4$ reduced by 1% (w/v) of sodium citrate, the mean diameter of AuNPs was found to be ~20 nm.[38] After the incubation, the fiber surface was covered with AuNPs, as was confirmed through the Scanning Electron Microscopy (S.E.M.) and Energy Dispersive Spectroscopy (E.D.S.) analyses explained later. Finally, the fiber was washed with DI water to remove any unattached nanoparticles from its surface.

**Experimental setup:** A fiber-coupled broadband light source (Part # SLS201L) from Thorlabs (Newton, NJ) and a FLAME-T-VIS-NIR-ES spectrometer from Ocean Insight (Orlando, FL) were used for carrying out the optical measurements. The surface functionalization was done on a $2 \times 1$ multimode fiber coupler (Part #TT200R5S1B) from Thorlabs, with one end connected to the light source and the other end connected to the spectrometer. The incident light propagated through the fiber via total internal reflection and illuminated the AuNPs-coated fused tip region. At a specific wavelength centered around 610 nm, the electrons within the AuNPs resonated with the incident light's electromagnetic field, resulting in localized surface plasmon resonance (LSPR). The spectrometer detected light reflected from the tip area. A dip in the reflection intensity was observed at LSPR. By tuning the size and shape of the nanoparticles, it is possible to control and manipulate the LSPR wavelength. In this work, the size and shape of AuNPs were adjusted to achieve the LSPR dip in the middle of the wavelength range (i.e., 350 – 900 nm) of the light source.[38]

**Diffusion of Fluid in Agarose-based Hydrogel:** An appropriate amount of agarose powder (Biotechnology Grade Agarose from V.W.R. life science, CAS:9012-36-6 and Lot: 21E0556711) was weighed on a balance (model: V.W.R. 314AC, S/N:650855) and mixed with DI water to make 1-5 % (w/v) solution of agarose. The solution was stirred on a hot/stir plate (V.W.R. 7x7" CER HOT/STIR 120 V pro, S.N.: 200929002) at 125 $^0$C and at a minimum of 100 rpm for approximately

15 minutes or until the agarose got dissolved completely. Approximately 8 mL of the solution was poured in a 10 mL small beaker, and then kept until it solidified. Methylene blue, M.B. (BioPharm, Hatfield, AR; 1% w/v, SKU: BM4292B) dye solutions were prepared by dilution with deionized water. A NanoJet pump (Chemyx Inc., S.N.: 48011) with PTFE Luer Lock 1 mL glass syringe (Hamilton Syringe) was used to control the fluid flow and deliver dye at a 2.5 µL/min rate for 60 minutes into the agarose matrix.

**Surface Characterization:** To analyze the surface coverage of the optical fiber with AuNPs, we used a scanning electron microscope (S.E.M.) (model-JSM-7100FT FESEM). The fiber surface was cleaned to eliminate any dust or debris by exposing the fiber to oxygen plasma treatment (PE-25V-HF) purchased from Plasma Etch. Due to the presence of a conductive material (i.e., AuNPs) on the sample surface, no additional sputter coating was required, and the fibers were directly placed on S.E.M. sample holders using double-sided carbon tapes.

*In Silico* **Models:** We simulated and used COMSOL Multiphysics version 6.1 for the *in silico* profile. We used the Free and Porous media flow and the transport of diluted species physics for the simulation. We declared a similar geometry of a 10 mL small beaker in 2D geometry, having 26.6 mm in height and 26 mm in diameter. The glass thickness was assumed to be 0.6 mm. The input catheter height was 13 mm, the inner diameter was 0.33 mm, and the outer diameter was 0.82 mm. We kept the flow rate and time the same as the *in vitro* diffusion studies, which we explained in the earlier section. The agarose properties and the fluid diffusivity were used the same as our previous work.[28]

## Results and Discussion

We first fabricated our LSPR optical fiber for the analysis and conducted the control study in a M.B. solution for calibration. Initially, we hypothesized to achieve a smooth signal from the light-matter interaction from our probe. Due to the slight variation in the interaction, we characterized the surface with scanning electron microscopy and investigated the topography and distributions of the AuNP on the optical fiber. After that, we analyzed the *in situ* concentration of the M.B. in various concentrations of agarose gel matrix at different heights representing the convection-enhanced drug delivery method and compared the results with *in silico* concentration profiles. The following sections provide more details on the results.

**Interaction of the Light with Methylene Blue in DI Water:** Prior to using the fiber-optic probe in the actual diffusive scenario within the agarose matrix, we aimed to assess the probe's sensitivity to varying concentrations of M.B. in DI water, as shown in **Figure 2**. Interaction with methylene blue is the control, presented as the percentage in reflection calculated using equation (1). We collected the optical response of the background before starting the data acquisition. Then, we collected the response of the bare fiber before the AuNP coating and dipped our AuNP-coated optical fiber in the sample. We calculated the reflection percentage according to the equation 1.

$$Reflection\ (\%) = \frac{AuNP coated\ fiber\ response - Background\ data}{Bare\ fiber\ response - Background\ data} X100\% \ldots\ldots\ldots (1)$$

**Figure 2a** shows the reflection spectra of the fiber-optic probe when it was immersed in five different concentrations of M.B. (i.e., 6.25%, 12.5%, 25%, 50%, and 100%) in DI water. Here, 100% M.B. solution corresponds to a concentration of 1.25 mol/m$^3$. We diluted M.B. with the DI water to achieve the desired concentrations. Optical reflection data in air and DI water (i.e., 0% M.B. solution) were also collected as the control, as shown in the black and red solid lines in **Figure 2a**. With increasing concentration of M.B., from 0% M.B. (i.e., only DI water) to 100% M.B., the reflection intensity decreased in **Figure 2a**. This phenomenon could be attributed to the increased scattering occurring at the LSPR surface upon interaction with a larger number of M.B. molecules. A similar phenomenon was also observed in our previous study.[30] The experiment was repeated with four identical fiber probes, exhibiting repeatability with a coefficient of variance of less than 1%. **Figure 2b** displays the calibration curve generated by plotting the mean reflected light intensity of four identical fiber-optic probes at the LSPR wavelength of around 610 nm against the M.B. concentration. The error bars represent the standard deviation in reflection intensities among the four different probes. The calibration curve can be used to estimate the unknown concentration of M.B. from the shift in LSPR intensity. It is well-established in the literature that the performance of the LSPR sensor is directly related to the size, shape, and density of metal nanoparticles.[39] LSPR sensors can also experience reduced signal-to-noise ratio due to nanoparticle aggregation if the processing temperature and time are not precisely controlled.[40] Therefore, we wanted to investigate the AuNP coating (distribution and topography) and characterize the surface of the optical fiber in the next section.

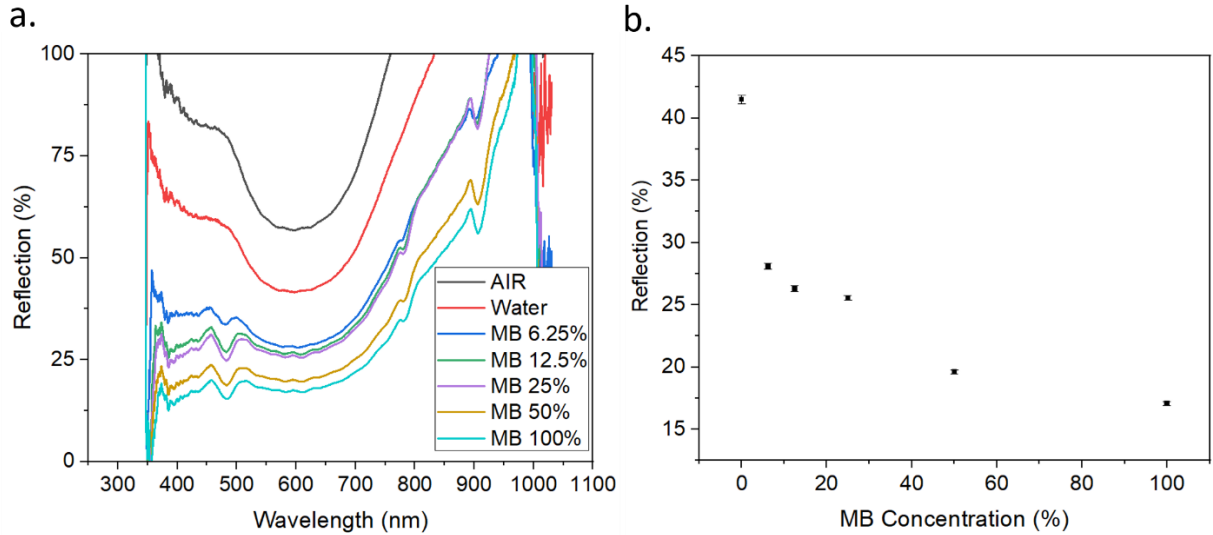

**Figure 2:** (a) Reflection spectra of the fiber-optic probe when it was immersed in different M.B. concentrations. (b) Calibration curve showing reflection variations among four identical probes.

**Surface Characterization of the Fiber:** The AuNP-coated optical fibers were analyzed using a scanning electron microscope (S.E.M.), having energy dispersive spectral (E.D.S.) capabilities to investigate the topography and distribution of the AuNP coating. The surface topography of an uncoated and an AuNP-coated fibers are shown in **Figure 3a** and **Figure 3b**, respectively. **Figure 3a** shows a smooth surface in contrast to the AuNP-coated fiber shown in **Figure 3b**. A closer examination of the AuNPs deposition on the silica fiber's surface was carried out, as shown in the magnified view of the coating in **Figure 3c**. The AuNP deposition was not entirely homogeneous according to our initial expectations. We noticed heterogeneity throughout the coating. These heterogeneities in the deposition may cause deviations and induce errors in the analysis. However, the impact of such deviation would be minimal since we focus on a very specific (i.e., 610nm) wavelength, and the heterogeneity-induced error would be negligible. E.D.S. analysis was conducted to identify the elemental composition in the region shown in **Figure 3d**, and the spectrum is presented in **Figure 3e**. AuNP resulted in the highest amount of detected Gold (Au) apart from Carbon (C), Oxygen (O), Silicon (Si), and Sulphur (S). Au, S, and C were detected due to the surface functionalization, and Si and O are the elements of the fiber-coating. The S was present because of the thiol groups (-S.H.) provided by MPTES. We further conducted the E.D.S. mapping of the whole region to investigate the distribution of these elements in **Figures 3f-h**. The different detected elements were represented with different colors in multiple images of the same surface. We can observe the Si and O distributed along the surface, where AuNP was absent. Au was presented in the most where AuNPs were attached to the surface. By employing basic image analysis techniques for area measurement with ImageJ,[41] we have quantified and distinguished the regions that received the coating and those that remained uncovered or uncoated. We have done such image analysis for **Figure 3d** and found that we achieved an AuNP coating of ~417.76 $\mu m^2$ among the total area of ~574.24 $\mu m^2$.

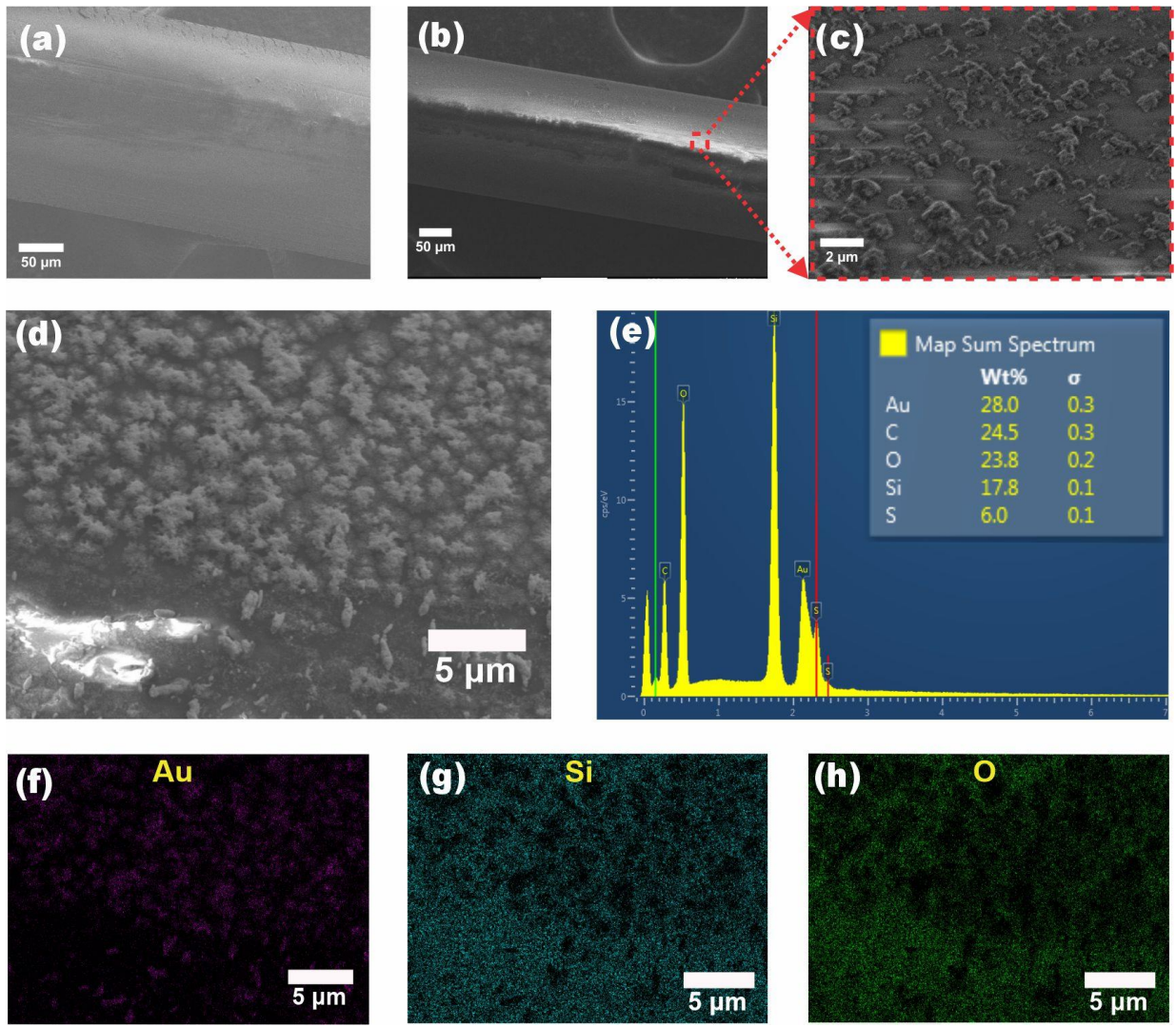

**Figure 3:** (a) The surface topography of uncoated optical fiber core. (b) The surface topography of AuNP-coated optical fiber core, with (c) showing the magnified topography in the S.E.M. image. (d) A representative area in the AuNP-coated fiber, (e) E.D.S. analysis on the image area[d]. Surface map of different elements, i.e., (f) Au, (g) Si, (h) O present in the coated fiber probe.

**Interaction of the Light with Methylene Blue in Gel Matrix:** We inserted the fiber-optic probe into a gel matrix and then introduced 1.25 mol/m$^3$ or 100% Methylene Blue, which was prepared in DI water, to examine its interactions with the fiber. A NanoJet pump was used to control the fluid flow and deliver the M.B. dye into the agarose matrix. The flow rate remained constant at 2.5 µL/min for 60 minutes. Therefore, a total of 150 µL of M.B. solution was infused into the gel matrix for each scenario. Three different concentrations (1%, 2%, and 5%) of gel were used for this experiment. The fiber was inserted at different Z-depths inside the gel (represented approximately in **Figure 4a**), with the resulting reflection spectra shown in **Figures 4b-d**. Here, Z1 represents a point above the diffusing area, and Z4 represents a point below the diffusing area. Z2 is slightly above the input tip into the main diffusing area, and Z3 is slightly below the input tip. We did not control the position strictly for simplicity.

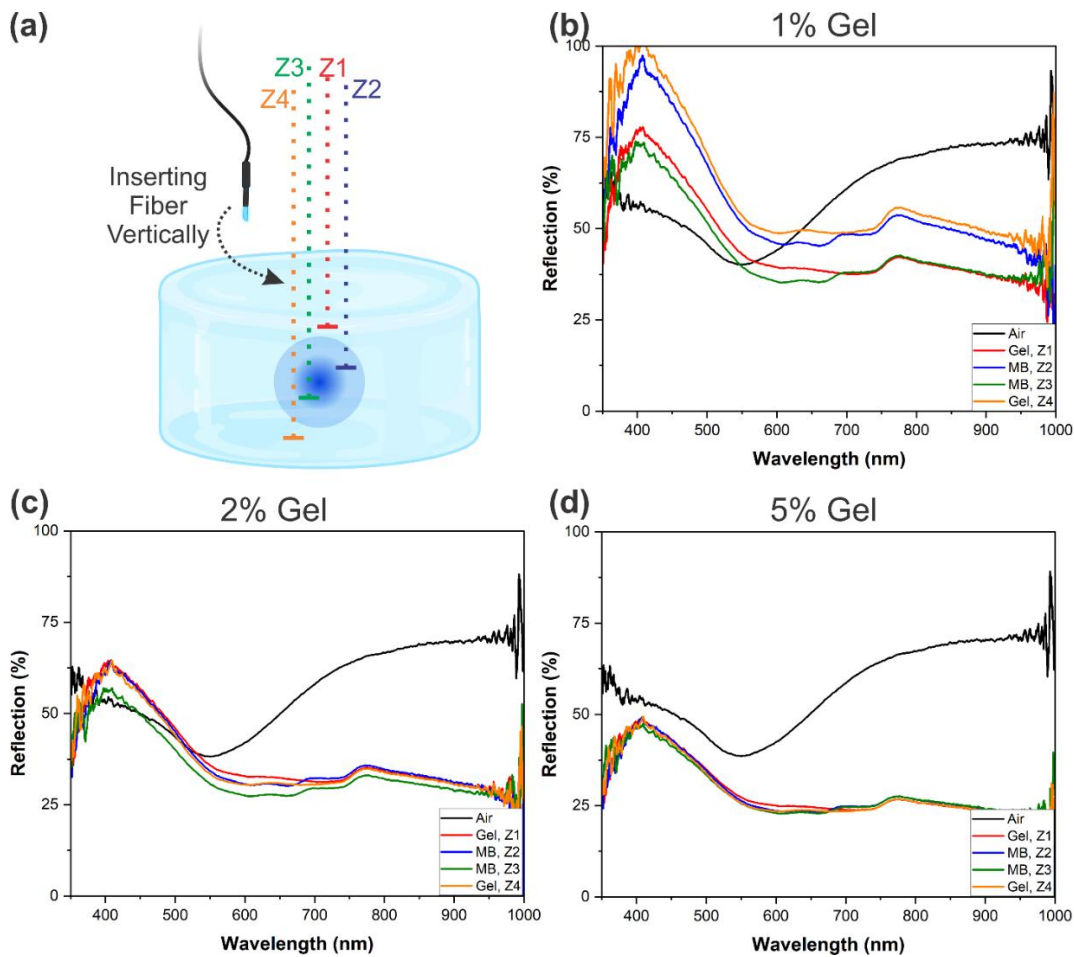

**Figure 4:** (a) Fiber probe inserted inside a gel matrix at varying Z heights. Reflection spectra for (b) 1% gel, (c) 2% gel, and (d) 5% gel.

For 1% gel, there was a noticeable shift in the reflection spectrum with the variations in the Z height. However, as the gel density increased, the reflection intensity started to drop significantly, making it difficult to differentiate the reflection spectra for Z height variation. The interaction from the agarose gel can be subtracted from the signal to measure the proper interaction with M.B. dye. We also note that agarose concentrations ranging between 0.6%-2% are used for most of the *in vitro* studies in literature.[24,27,28] As we observe significant interactions up to 2% agarose gel, our fiber optic system is expected to probe the local concentrations without facing challenges. Additionally, we tested the reproducibility of the fiber probe and presented the result in **Figure S1**. The measurement was repeated for the same Z height in the same gel concentration. Although a small offset in the reflection spectrum is observed, it can be easily corrected by incorporating a constant bias to the measured intensity value.

***In Silico* Comparison and Validation:** We can measure the local concentration at variable positions with our presented optical fiber. To extract the complete concentration profile, we can combine *in vitro* measured data with our already developed and previously presented *in silico* models.[28] We simulated the diffusive scenario in an agarose matrix and extracted the line concentration profile at different horizontal distances. As we insert the optical fiber vertically, we also define the cut line vertically at different horizontal distances. The complete concentration profile can be extracted from the simulation software by inputting the exact distance used for the concentration measurement in the *in vitro* scenario. If we know the exact horizontal distance of the optical fiber from the input catheter, we can extract the *in silico* concentration profile and compare the input height with the vertical distance. The simulated 2D geometry is presented in **Figure 5a**. We used the properties of 1% agarose gel and diffused M.B. at a similar flow rate and investigated the concentration profile. The concentration profile after 60 min is also presented in

**Figure S2**. We defined multiple cut lines at different horizontal distances from the center of the input catheter and extracted the vertical concentration profiles. Different vertical concentration profiles are presented in **Figure 5b**. We presented the concentration in percentages, where 100% M.B. concentration represents 1.25 mol/m$^3$. The negative distance indicated the left side of the input catheter. At a 9 mm distance, we can see a very insignificant concentration, which is far from the input catheter and the concentration front. We also can observe a similar profile on the right and left side of the catheter, as the diffusion profile follows a similar error function on both sides. At a 1 mm distance, the concentration was observed 100% up to ~7mm, as concentration is supposed to reach saturation near the input flow. Thus, the optical fiber response can be harnessed with the *in silico* profile to validate and asses the molecular distribution of analytes.

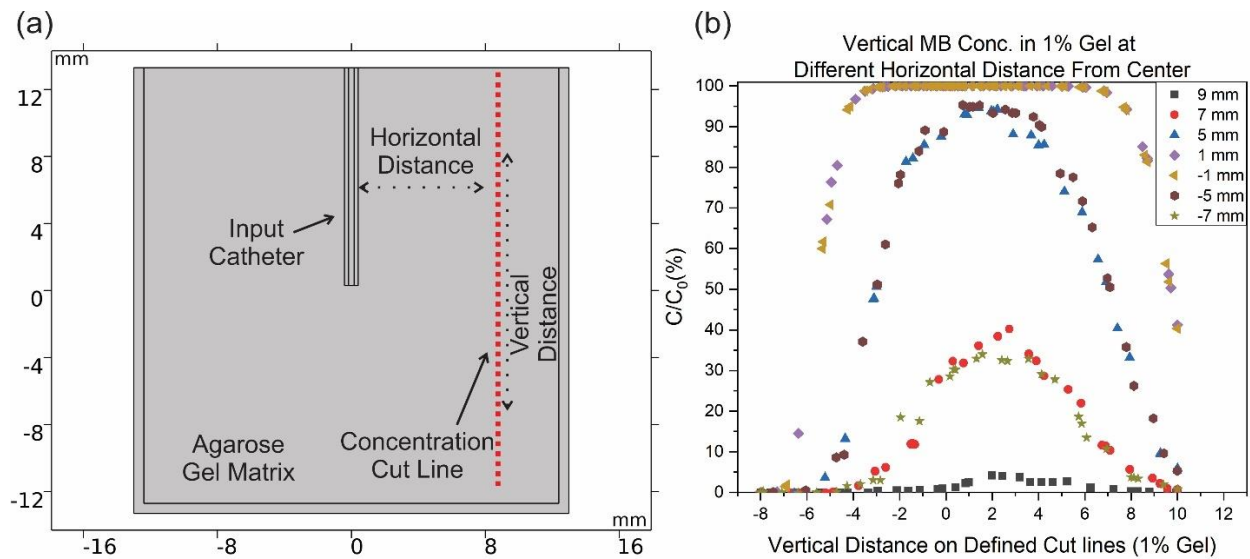

**Figure 5:** (a) *in silico* models to simulate and extract the complete concentration profile, (b) extracted vertical M.B. concentration profiles in 1% agarose gel at different cut lines, defined at different horizontal distances from the input catheter.

## Conclusion:

In this study, we have shown the feasibility of probing diffused analyte in a complex biomatrix, leveraging AuNP-coated optical fiber. While expensive analytical tools lack such simplicity, our platform demonstrates a proof of concept. We have presented a facile system to investigate the *in situ* concentration of a synthetic drug within an *in vitro* system. We have diffused methylene blue dye in different concentrations of agarose gel matrix and investigated the local *in situ* concentrations with an LSPR-based optical fiber, which was fabricated without using any cleanroom or expensive steps. We further complemented the *in vitro* concentration findings with our already developed and previously presented [28] *in silico* models to understand the molecular distribution of the synthetic drug. Such a simple method of investigating the localized distribution of analytes will open numerous opportunities to investigate the *in situ* concentrations of different biomarkers *in vivo* or *in vitro* systems and improve the overall treatment technology.

## Author Contributions:

## Acknowledgement:

We acknowledge Lawrence Ray, a former M.S. student at the University of Nevada, Reno, for initially assisting with the experimental procedures and Dr. Joel DesOrmeau for his assistance during the SEM-EDS analysis. We acknowledge the funding support from the University of Nevada, Reno-VPRI startup fund, and partial support from the National Institute of General Medical Sciences of the National Institutes of Health under grant number P30 GM145646. M. R. K. acknowledges the partial funding support for this research provided under a cooperative agreement with the U. S. Army Corps of Engineers, Engineer Research and Development Center, Construction Engineering Research Laboratory (USACE ERDC-CERL), under the contract W91T32T-23-2-0004 (Advanced Technologies and Systems for Water Reuse, 2023) through the Nevada Center for Water Resiliency (NCWIR). S. T. acknowledges the funding support from the National Science Foundation under award number ECCS 2138701 and the VentureWell under award number 21716-20.

# Gold Nanoparticles Coated Optical Fiber for Real-time Localized Surface Plasmon Resonance Analysis of *In-situ* Light-Matter Interactions

Nafize Ishtiaque Hossain[1,*], Kazi Zihan Hossain[2,*], Momena Monwar[2], Md. Shihabuzzaman Apon[3,┴], Caleb Shaw[2,┴], Shoeb Ahmed[3], Shawana Tabassum[1,#], M. Rashed Khan[2,#].

[1]Department of Electrical Engineering, University of Texas, Tyler; Texas, U.S.A.

[2]Department of Chemical and Materials Engineering, University of Nevada, Reno; Nevada, U.S.A.

[3]Department of Chemical Engineering, Bangladesh University of Engineering and Technology; Dhaka, Bangladesh.

*,┴Contributed equally

[#]Corresponding Authors: Shawana Tabassum: stabassum@uttyler.edu

M. Rashed Khan mrkhan@unr.edu

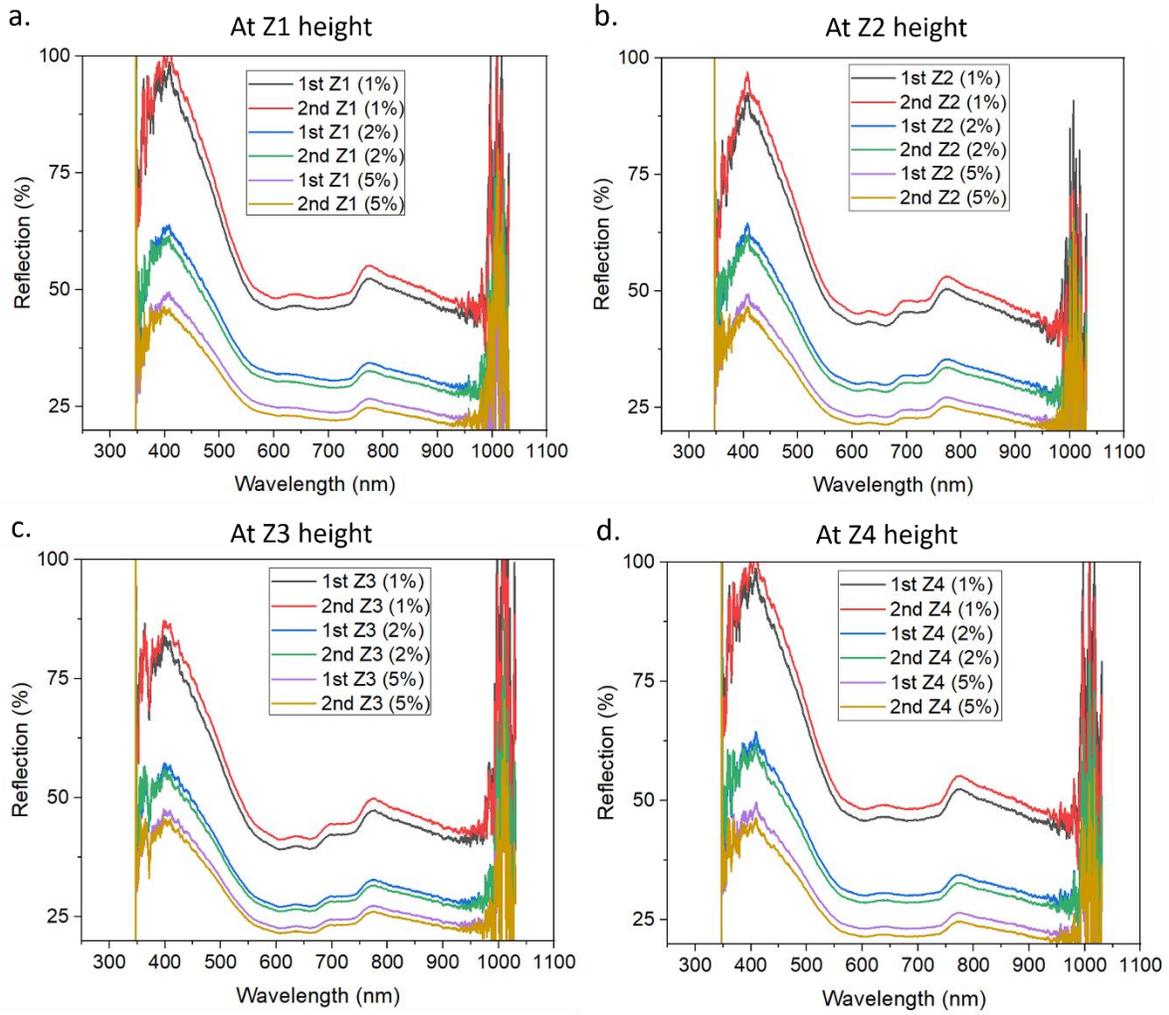

**Figure S1:** Reproducibility test of the fiber probe when measurement was repeated for (a) Z1, (b) Z2, (c) Z3, and (d) Z4 heights.

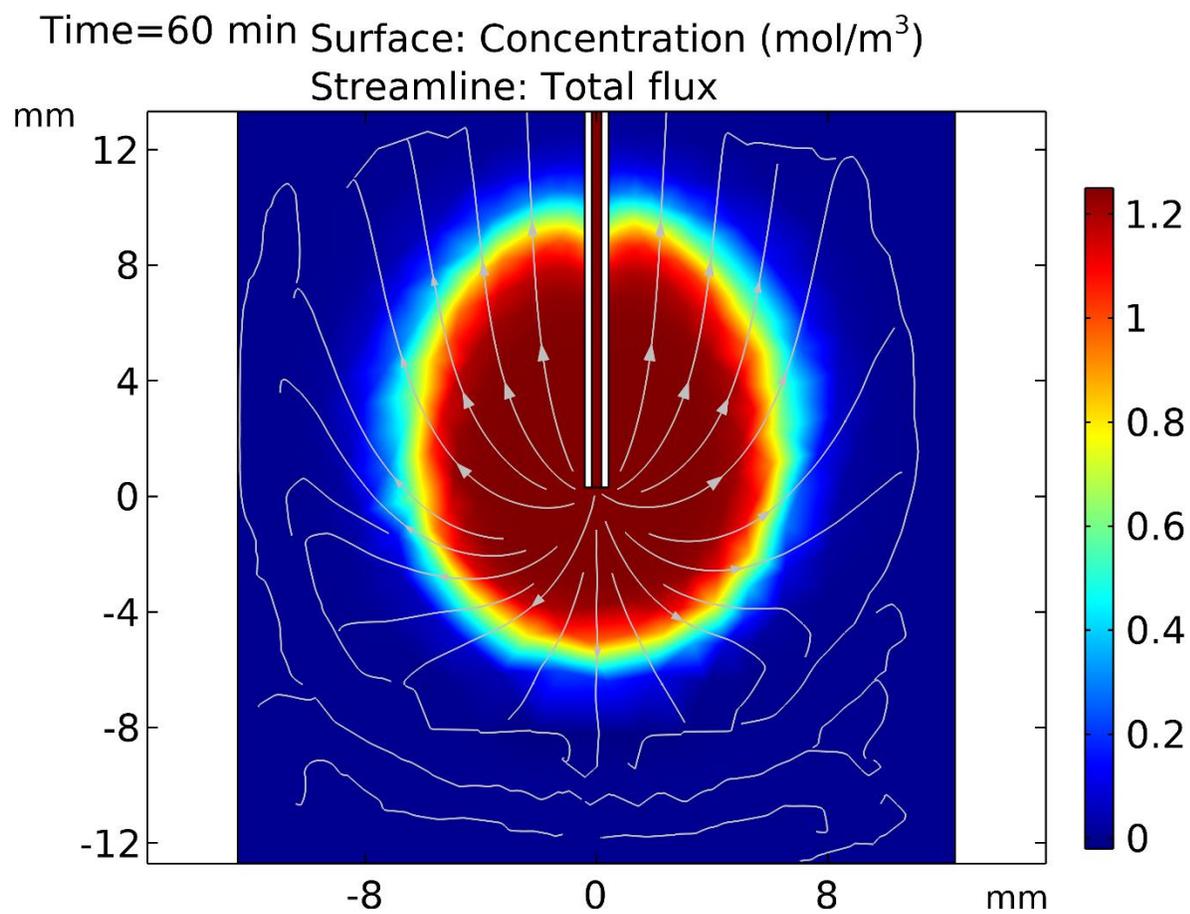

**Figure S2:** *In silico* concentration profile of MB in 1% agarose gel after 60 min.